\documentclass[runningheads]{llncs}
\usepackage[T1]{fontenc}
\usepackage{graphicx}
\usepackage{multirow}
\begin{document}
\title{AutoIE: An Automated Framework for Information Extraction from Scientific Literature}
%
%
\author{Yangyang Liu\inst{1}\orcidID{0009-0001-0539-338X} \and
Shoubin Li\inst{2,*}\orcidID{0000-0001-8641-8130} }
%
%
\institute{University of Auckland, Auckland, New Zealand
\email{yilu660@aucklanduni.ac.nz}
\and
Institute of Software Chinese Academy of Sciences, Beijing, China
\email{shoubin@iscas.ac.cn}}
\maketitle              
\begin{abstract}
In the rapidly evolving field of scientific research, efficiently extracting key information from the burgeoning volume of scientific papers remains a formidable challenge. This paper introduces an innovative framework designed to automate the extraction of vital data from scientific PDF documents, enabling researchers to discern future research trajectories more readily. AutoIE uniquely integrates four novel components: (1) A multi-semantic feature fusion-based approach for PDF document layout analysis; (2) Advanced functional block recognition in scientific texts; (3) A synergistic technique for extracting and correlating information on molecular sieve synthesis; (4) An online learning paradigm tailored for molecular sieve literature. Our SBERT model achieves high Marco F1 scores of 87.19 and 89.65 on CoNLL04 and ADE datasets. In addition, a practical application of AutoIE in the petrochemical molecular sieve synthesis domain demonstrates its efficacy, evidenced by an impressive 78\% accuracy rate. This research paves the way for enhanced data management and interpretation in molecular sieve synthesis. It is a valuable asset for seasoned experts and newcomers in this specialized field.

\end{abstract}
\section{Introduction}
Recent statistics reveal a general trend of an approximate 4\% annual increase in the publication rate of scientific and technological papers, signifying the rapid advancement and dynamic nature of contemporary research ~\cite{cite_1}. The burgeoning volume of scientific literature precipitates a critical challenge: the urgent need for efficient and precise automated information extraction technologies.   Traditional methods, once reliable, now struggle to cope with data's expanding scale and complexity, rendering them increasingly ineffective. Our study proposes an innovative framework, AutoIE, for rapidly and accurately extracting essential information from scientific texts. 

We identify several key obstacles in developing a robust framework for this purpose:
\begin{itemize}
\item (1)  \textbf{Length and Complexity}: The length of scientific and technological papers is very long, and it is difficult to quickly locate and mine the valuable information in the text in such a long length. For example, to extract the comparative results of various experimental methods in scientific and technological papers, it is only necessary to extract the "experimental results" chapter of the paper. In this process, a large amount of text and data information in other chapters are noise data;
\item (2)  \textbf{Limitations of Current Information Extraction Methods}: The current information extraction methods have achieved good results in the open field, but these technologies can not be well transferred to the key information extraction field of academic papers;
\item (3)  \textbf{Domain-Specific Challenges}: Scientific literature is characterized by dense domain knowledge and high specialization.   Creating a domain-specific sample database involves intricate processes requiring expert involvement, low automation levels, and stringent quality demands.   These factors hinder the adequate labelling of domain datasets, limiting the training of deep learning-based models for efficient entity extraction in specialized domains.
\end{itemize}

To address these challenges, we propose the AutoIE (Framework architecture is shown in Fig. \ref{fig1}), which consists of three innovative components (Layout and Location Unit, Information Extraction Unit and Display and Human Feedback Unit).
In the Layout and Location Unit, we use the Multi-Semantic Feature Fusion-based Approach for PDF Document Layout Analysis (MFFAPD) and Advanced Functional Block Recognition in Scientific Texts (AFBRSC) model to quickly locate the location of the information to be extracted in the document. In the Information Extraction Unit, we propose a new entity and relationship joint extracting model, SBERT, and combine it with transfer learning to extract the key information from the specific field- scientific documentation. Finally, in the Human Feedback Unit, we use the Online Learning Paradigm Tailored Method (OLPTM) to address challenges of corpus scarcity, expert availability, and sample labelling in specialized domains. Furthermore, to demonstrate AutoIE's efficacy, we applied it in the molecular sieve domain, yielding promising results.

Our contributions are summarized as follows:
\begin{itemize}
\item (1) \textbf{SBERT's Technical Superiority}: Our study highlights the technical superiority of the SBERT model, which not only achieved high Marco F1 scores of 87.19 and 89.65 on CoNLL04 and ADE datasets but also exhibited remarkable accuracy of 78\% in the specialized task of extracting key information from scientific literature in molecular sieve synthesis. These results are indicative of SBERT's advanced capabilities in complex data environments.
\item (2) \textbf{AutoIE}: The research introduces an automated framework, AutoIE, specifically engineered to overcome the challenges in key information extraction from the scientific literature. Our study provides a new paradigm for information extraction from scientific-technological literature. Furthermore, we also demonstrate the effectiveness of the AutoIE in Molecular sieve synthesis scientific documents.
\end{itemize}
\section{Related Work}
\subsection{Layout Analysis}
With the development of deep learning, there has been a paradigm shift in layout analysis techniques. Deep learning models offer enhanced characterization and modelling capabilities, handling complex detection tasks more efficiently. Notable models include R-FCN ~\cite{cite_21}: An approach optimizing region-based detection, offering improvements in processing efficiency. SSD ~\cite{cite_22} and YOLO ~\cite{cite_23}: Both models made significant strides in real-time processing, which is crucial for large-scale document analysis.
The novelty of this method lies in its ability to integrate and process multiple types of data for more accurate layout recognition. Another significant contribution in 2017 ~\cite{cite_24} involved a deep learning model that dramatically improved the accuracy of scientific document layout analysis. However, these methods have some problems in identifying complex, lengthy, and domain-specific scientific and technical paper layouts.

\subsection{Named Entity Recognition} 
The joint extraction model based on shared parameters will share input features or internal hidden layer states, and the loss value generated by entity recognition will be added to the loss value generated by relation classification. Bowen et al. ~\cite{cite_28} used pointer networks to decode entities, and the model proposed a new decomposition strategy, which hierarchically decomposes the task into several sequence labelling problems. Zeng et al. ~\cite{cite_29} used the Encoder-Decoder mechanism to obtain possible relationships in the sequence and then used the copy mechanism to extract potential head and tail entities from the input sequence simultaneously. Eberts et al.  ~\cite{cite_6} used the transformer network and combined the span method.  ~\cite{cite_30} Compared with  ~\cite{cite_6}, it is slightly different.  After extracting all entities, the relation and the corresponding head entity are output simultaneously, eliminating the need to construct entity pairs.

Using BERT to construct span vectors has achieved the state-of-the-art effect in the joint extraction model, but few researchers combine part-of-speech features with BERT’s word embedding features. Fatema et al. ~\cite{cite_34} studied the problem of combining Bert embedding with traditional NLP features in relation extraction. However, to our knowledge, there is no research on the effectiveness of combining part-of-speech features with BERT context embedding for the joint extraction model. Our SBERT model combines part-of-speech features with word embeddings generated by BERT and uses a joint extraction model based on the span method to extract entities and relations.

\section{Framework}
\begin{figure}
\includegraphics[width=0.9\textwidth]{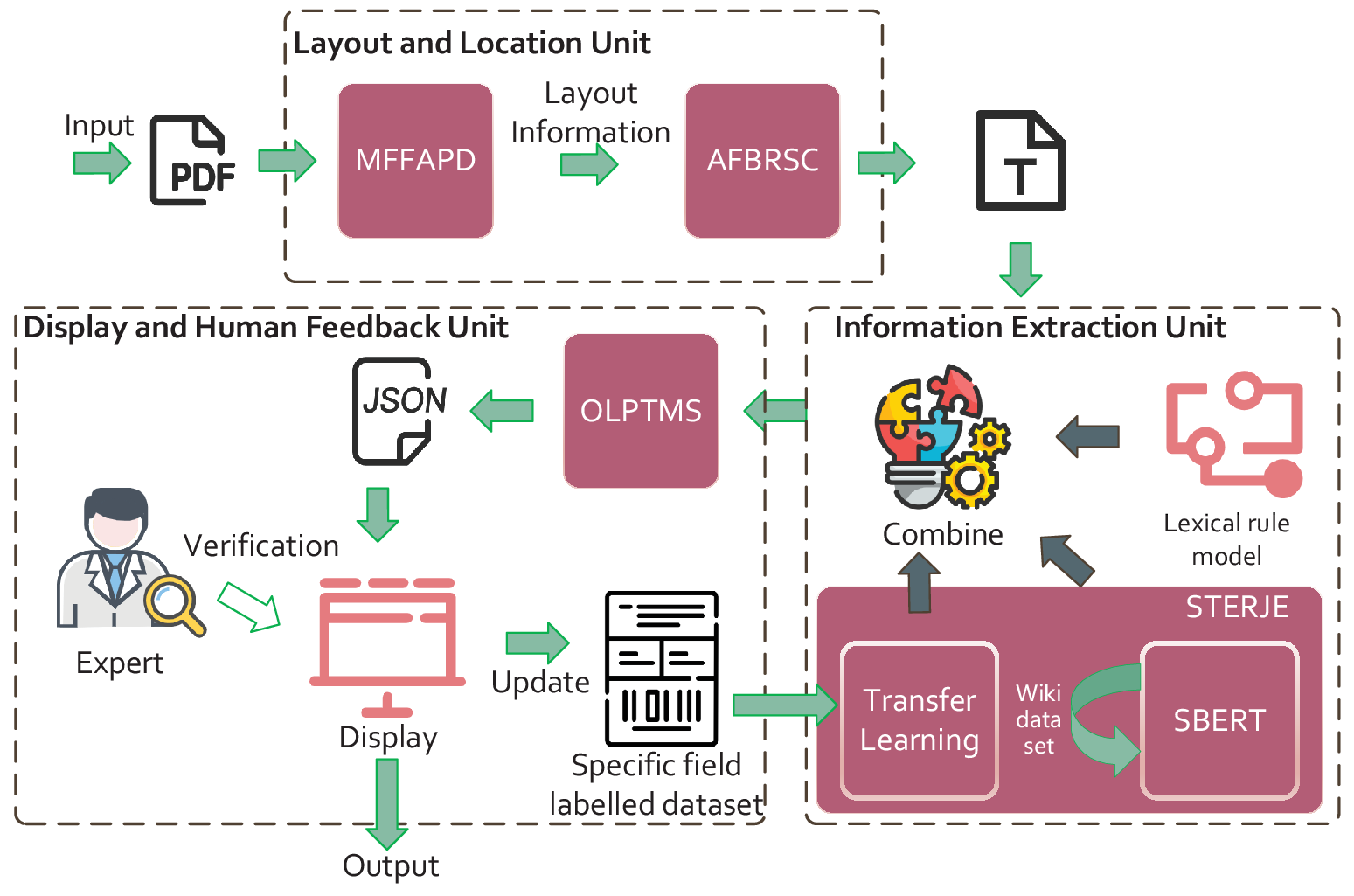}
\caption{High Level of AutoIE.} \label{fig1}
\end{figure}

In this section, we describe the framework architecture of our framework. We explain our approaches and algorithms in detail. We also show the process flow of our framework (shown in Fig. \ref{fig1}), which will, according to the flow, introduce every part of the framework.

\subsection{AutoIE Work-Flow}
\subsubsection{3.1.1 Layout and Location Unit}
\
\newline
\indent
The Layout and Location Unit takes PDF documents as input and gives layout division and generic of the document as output for the Information Extraction Unit. It employs a two-pronged approach for analyzing and locating the structure of PDF documents. Firstly, we use VTLayout ~\cite{cite_2} to recognize each block of the PDF document, such as the title, paragraph, table, figure, formula, etc. This method combines the low and deep layer visual and text semantic features. Therefore, it has been well-represented in the scientific literature and has done an excellent job of coarse-grained block localization for subsequent information extraction.

In addition, the most crucial composition information of molecular sieves is located in the method and experiment part of the paper according to the experience. Hence, we use HARGSD ~\cite{cite_3} to locate quickly and accurately. HARGSD dramatically reduces the influence of noisy data and improves the speed and accuracy of information extraction.
\subsubsection{3.1.2 Information Extraction Unit}
\
\newline
\indent
The Information Extraction Unit is a crucial component in our framework, processing layout and generic information received from the Layout and Location Unit. Its primary function is to generate pre-labelled data, which the Display and Human Feedback Unit subsequently utilizes. Initially, the unit employs the acquired layout information to accurately reconstruct the textual data in its correct sequence, saving it in a text format. This text data and location information are inputted into the Sentence-BERT (SBERT) model for advanced semantic processing. The outcome of this step is a temporary JSON file, meticulously organized to store the extracted key information, such as ${\{content: "", "labels": [], "connections": []\}}$.

\subsubsection{3.1.3 Display and Human Feedback Unit}
\
\newline
\indent
The Display and Human Feedback Unit is integral to our research framework, primarily handling the temporary JSON files generated by the Information Extraction Unit. Our process involves a meticulous review by our annotation team, who access the temporary JSON data via a dedicated web page system. They rigorously check the accuracy of the pre-labeled data, making necessary amendments to ensure high-quality inputs for model training. Once the data passes this human verification stage, it is systematically stored in our database. This refined data is crucial for enhancing the SBERT model's learning process, ensuring its efficacy in handling complex information extraction tasks. This unit's outputs play a vital role in the continuous improvement of our framework, linking closely with both the Information Extraction and Model Training Units to advance our research objectives.
\subsection{Method} 
In this section, we deploy the SBERT model, a state-of-the-art approach for key information extraction from scientific PDF documents. The model operates on a multi-dimensional feature analysis framework, which is crucial for accurate semantic interpretation.

\begin{figure}
    \centering
    \includegraphics[width=0.9\textwidth]{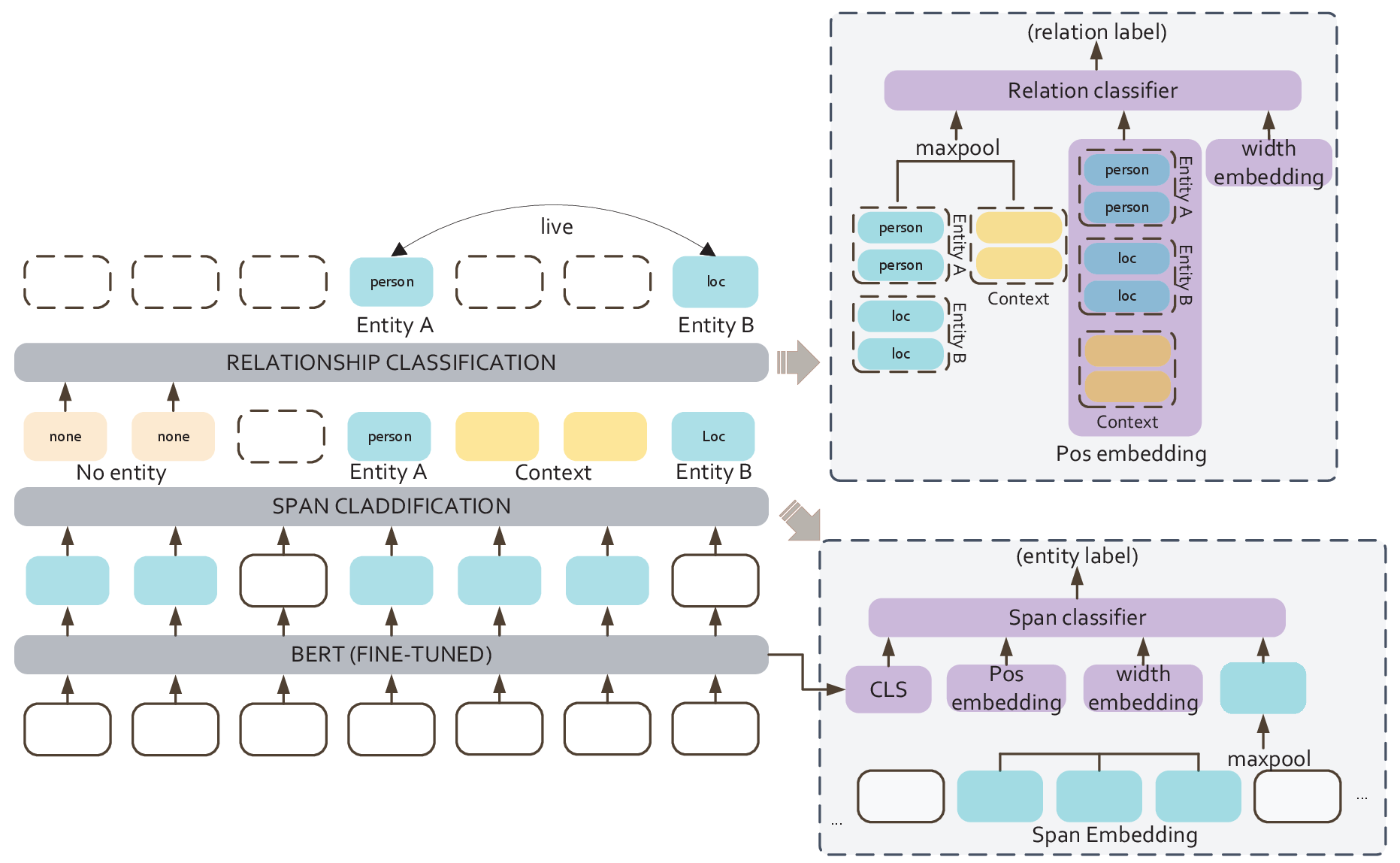}
    \caption{The architecture of SBERT.} 
    \label{fig2}
\end{figure}

\subsubsection{3.2.1   Multi-dimensional features}
\
\newline
\indent \textbf{Span Embedding}: This phase involves tokenizing sentences (use spacy tool ~\cite{cite_4} to split input sequence $ S = \omega _1  \omega _2 ...\omega_n$ into token sequences $[t_1,t_2,...,t_n]$ and generating span sequences. Each span is encoded by a fine-tuned BERT vocabulary, with the final layer's hidden state as the span embedding. This method, proven effective in semantic feature extraction, ensures the nuanced capture of contextual information.

\textbf{Width Embedding}: Span width conversion into a vector representation is a pivotal step ~\cite{cite_5}, filtering out excessively long or irrelevant span entities. This embedding process learned through backpropagation, enables the model to incorporate context-specific priors effectively ~\cite{cite_6}.

\textbf{CLS}: Utilizing the BERT model's CLS marker ~\cite{cite_7}, we capture the comprehensive semantic representation of entire sentences. This feature is especially beneficial for downstream classification tasks, enhancing our model’s predictive accuracy.

\textbf{POS Embedding}: To encode part-of-speech features, we use NLTK ~\cite{cite_8} to convert tokens into part-of-speech symbols. We use binary coding to design part-of-speech vectors. The part-of-speech type in the NLTK package is fixed, and the maximum part-of-speech dimension is designed according to the total number of parts of speech, thereby constructing a part-of-speech index table. The POS embedding of each token is obtained by querying the index table. This methodology allows for retaining relative positional information within the span, adding a syntactic layer to our analysis.
\subsubsection{3.2.2 Model Architecture}
\
\newline
\indent Our model's architecture (shown in Fig. \ref{fig2}) is tailored for scientific text analysis. It encompasses two principal components: span classification and relational analysis. Initially, the model classifies spans to pinpoint key entities. Subsequently, it examines the relationships between these entities, contextualizing them within the broader scientific discourse.

\textbf{Span classification}: The internal architecture of this stage is shown in the lower right half of Fig. \ref{fig2}. The input is a span composed of tokens, and the output is the entity tag of the span. The number of entity tags is k+1 (the predefined k entity types plus one none type). Span vector is represented by $S = (t_1 , t_2, ... , t_n)$. This part is transformed into $f{(t_1 , t_2, ... , t_n)}$. after max-pooling function. In ~\cite{cite_6}, it is verified through experiments that max-pooling is more effective than other pooling operations regarding text feature representation. Therefore, we also use the max-pooling method to extract the maximum feature of the span vector. The span before BERT encoding is the original text, represented by $W = (\omega _1, \omega _2, ..., \omega_n)$. After each token in the span is converted into part of speech, the part of speech vector $p{(\omega _1, \omega _2, ..., \omega_n)}$ can be obtained by binary coding. To filter entities whose span length is too long, we add the width feature of the span. The length of the input span is $k$, and the width embedding obtained after looking up the word embedding table is $E_k$. Therefore, what is finally entered in the span classifier is:

\begin{equation}
x^{s} =f{(t_1 , t_2, ... , t_n)}*p{(\omega _1 , \omega _2, ..., \omega_n)}*E_k*c     
\end{equation}

Where $c$ means CLS classification mark, and $*$ means vector splicing. The spliced vector passes through the softmax classifier, and each span is labelled with an entity label (including the
none label).

\textbf{Relation classification}: The internal architecture of this stage is shown in the top right half of Fig. \ref{fig2}. After screening the spans classified as none, $N$  span entities remain. The generated candidate entity pairs have N*N groups. The model's input is N*N entity pairs, and the output is the relationship label of each entity pair. The characteristics of each entity pair input to the relational classifier are composed of three categories:
\begin{itemize}
    \item 1) Let $(e_1, e_2)$ denote the input entity pair vector, and $c{(e_1, e_2)}$ denote the text encoding between the entity pairs. According to experiments ~\cite{cite_6}, it is proved that the text between entities is compared with the whole sentence and CLS label. Regarding the relationship classification effect, the text between entity pairs performs best. Therefore, we use the text between entity pairs as the relation classification feature. The entity pair and the text pass through the max-pooling layer respectively, and the generated vectors are represented as $f{(e_1, e_2)}$  and $f{(c{(e_1, e_2)})}$.
    \item 2) Considering the importance of the part-of-speech order at the junction between the head and tail entities and the text, the entity and the text are transformed into part-of-speech features. The entity pair before BERT encoding is represented as $(\omega_{e1}, \omega_{e2})$, and the text between entity pairs is represented as $c{(\omega_{e1}, \omega_{e2})}$. After being converted into a part-of-speech sequence, it is encoded in binary. Entity pair and text correspond to $p{(\omega_{e1}, \omega_{e2})}$ and $p{(c{(\omega_{e1}, \omega_{e2})})}$ respectively. After adding part-of-speech features, the model can filter out entity pairs that do not conform to the regular part-of-speech order in relation extraction.
    \item 3) The width embedding of entity pairs is also used as a feature of relation classification. The length of the token of the input entity pair is expressed as $(k_1, k_2)$, and the converted width embedding is expressed as $W{(\omega_{e1}, \omega_{e2} )}$. 
\end{itemize}

The three types of feature vectors are spliced and input into the relational classifier. The spliced vector is expressed as follows:
\begin{equation}
x^{r} =f{(e_1 , e_2)}* f{(c(e_1 , e_2))}*p{(\omega _{e1} , \omega _{e2})}*p{(c(\omega _{e1} , \omega _{e2}))}*(E_{k1}, E_{k2})
\end{equation}

Where $*$ means vector splicing. The spliced vector is input to the single-layer classifier and activated by the sigmoid function. The relationship label with the highest score in the sigmoid layer is taken as the relationship between the entity pairs. At the same time, the threshold is set, and the sigmoid function can only be activated if it is greater than the threshold. Otherwise, there is no relationship between the entities.

\section{Experiment}
This section describes data, experiments, and evaluations of our results. We first use two public datasets to evaluate the effectiveness of SBERT's performance in the general entity-recognized field. We then used our framework to extract key information from the scientific literature in molecular sieve synthesis to demonstrate that our framework can extract key information from the scientific literature quickly and accurately.
\subsection{Data}
Our model is evaluated on two publicly available data sets and one private dataset.
\begin{itemize}
    \item (1)  CoNLL04: The entity and relation annotations in the Conll04 data set ~\cite{cite_9} come from news reports. There are four types of entities: Location, Organization, People, and Other. There are five types of relationships: Work-for, Kill, Organization-Based-In, Live-In, and Located-In. Our training set and test set are the same as ~\cite{cite_10}; the training set has 1153 sentences, and the test set has 288 sentences.
    \item (2)  ADE: The ADE dataset ~\cite{cite_11} describes the impact of drug use in medical reports. The entity categories are Adverse-Effect and Drug, and the relationship category is only Adverse-Effect. We adopted ten verifications consistent with the work in ~\cite{cite_6}.
    \item (3) The molecular sieve synthesis dataset is a scientific PDF document (almost 2000 pdf) from the molecular sieve synthesis field. Key information, which is extracted from the document, is divided into two fields: public field and private field. The public field includes title, author, unit, Published time, magazine, DOI, and product zeolite. The private field includes Alkali Source, Cations, Fluorine Source, Crystallization Conditions temperature, Crystallization Conditions time, Crystallization Conditions rotational speed, Germanium Source, Gel Composition, Phosphorus Source, Silicon Source, Template Agent, Aluminum Source, Molecular Sieve Structure Information.
\end{itemize}

\subsection{Baseline and Parameter Set}

\textbf{Baseline}: As mentioned, we use two publicly available datasets to evaluate SBERT. To further highlight the validity of our model in the general field, we use a variety of complex neural arrest models as the baseline (multi-head+AT ~\cite{cite_12}, multi-head ~\cite{cite_13}, Relation-Metric with AT ~\cite{cite_14}, Relation-Metric with AT ~\cite{cite_15}, Spert ~\cite{cite_6}).  

\textbf{Model Configuration}: Pre-training Model: We utilized the BERTBASE (cased) model, which was pre-trained on an English corpus. Learning Rate and Dropout: The model's learning rate is 5e-5, with a dropout rate 0.1.

\textbf{Training Parameters}: Span Length and Width Embedding: The maximum span length is 10, and the width embedding dimension is 25. Epochs, Batch Size, and Relational Filtering: We configured the model to run for 30 epochs with a batch size of 2. The threshold for relational filtering is set at 0.4.   Negative Sampling: We capped the maximum negative samples for entities and relationships at 100 due to the predominance of negative samples from the span method. Part-of-Speech (POS) Features: Each POS is assigned a dimension of 6.      Considering the maximum span length of 10, the total dimension for each span's POS coding is fixed at 60. Spans shorter than ten are padded with zeros.

\textbf{Loss Calculation}: Entity Recognition and Relation Classification: The loss value during training is the sum of the loss from entity recognition and relation classification.      Entity recognition employs cross-entropy loss, while relation classification uses binary cross-entropy loss.

\textbf{Evaluation Metrics}: We assessed the model using F1 scores and extraction accuracy. In the following formula, the TP stands for True Positive, the TN stands for True Negative, the FN stands for False Negative, and the FP stands for False Positive

\begin{equation}
Accuracy = \frac{TP+TN}{TP+TN+FP+FN}     
\end{equation}
\begin{equation}
F1 = \frac{2TP}{2TP+FN+FP}    
\end{equation}
\textbf{Validation Method}: Each experiment employed a ten-fold cross-validation approach for training the model.

In our evaluation using the CoNLL04 dataset(Table \ref{tab1}), our model demonstrates superior performance in both entity recognition and relationship extraction tasks compared to the established baseline models. Specifically, in the entity recognition task, our model shows an improvement with a near 1\% increase in Macro F1 score, topping the benchmark for this dataset. Previously, the highest recorded baseline performance was an F1 score of 72.87. This improvement suggests that incorporating part-of-speech features significantly enhances the model's ability to recognize entities. However, while both entity recognition and relationship extraction show improvements over the baseline, the enhancement in entity recognition is more pronounced. This is likely due to the dataset's extended average sentence length characteristic, which presents more complex scenarios for capturing relationships between entity pairs using part-of-speech features. Our findings on the ADE dataset offer a different perspective. Here, the improvement in relationship extraction over the baseline is more significant than entity recognition.

\subsection{Result of SBERT}
\begin{table}
\centering
\caption{Ten fold cross verification results on CoNLL04 and ADE.}\label{tab1}
\begin{tabular}{|c|c|c|c|}
\hline
Dataset	& Model	& RE Marco F1	& NER Macro F1 \\
\hline

\multirow{6}{*}{CoNLL04}
        & Multi-head+AT	& $61.95$	& $83.60$\\
	& multi-head	& $62.04$	& $83.90$\\
	& Relation-Metric with AT	& $62.29$	& $84.15$\\
	& Biaffine attention	& $64.40$	& $86.20$\\
	& Spert	& $72.87$	& $86.25$\\
	& SBERT [Ours]	& \textbf{$73.18$}	& \textbf{$87.19$}\\
\hline
\multirow{6}{*}{ADE}
        & Multi-head+AT	& 74.58	& 86.40\\
	& multi-head	& 75.52	& 86.73\\
	& Relation-Metric with AT	& 77.19	& 87.02\\
	& Biaffine attention	& 78.84	& 89.28\\
	& Spert	& 79.24	& 89.25\\
	& SBERT [Ours]	& \textbf{79.84}	& \textbf{89.65}\\

\hline
\end{tabular}
\end{table}

The ADE dataset (Table \ref{tab1}), focusing on medical field data, contrasts with the CoNLL04 news report data, encompassing a broader range of professional terms. Consequently, the impact of adding part-of-speech features on entity recognition is less pronounced than in the CoNLL04 dataset. This difference can be attributed to the ADE dataset's shorter average sentence length, which simplifies capturing the part-of-speech patterns between entity pairs and within the text surrounding these pairs. Additionally, the sequence of part-of-speech tags at the junctions of entities and their contextual text appears to aid the model in more effectively identifying relationships between entity pairs.

\subsection{Application of AutoIE}
\
\newline
\indent Our experimental results demonstrate that the SBERT model, adapted to the molecular sieve synthesis domain through transfer learning, outperforms existing joint entity-relation extraction methods. This adaptation involves leveraging a multi-feature fusion-based joint extraction model, a novel approach in this specialized field.
\begin{table}
\centering
\caption{Key information extraction accuracy.}\label{tab2}
\begin{tabular}{|l|l|l|l|}
\hline
Field &  Accuracy & Field &  Accuracy \\
\hline
Title &  0.93 & Alkali Source &  0.54\\
Author &  0.96 & Cations & 0.60\\
Unit & 0.90 & Fluorine Source & 0.63\\
DOI & 0.95 & Crystallization Conditions temperature & 0.77\\
Published Time & 0.91 & Crystallization Conditions time & 0.80\\
Zeolite & 0.79 & Crystallization Conditions rotational speed & 0.85\\
Magazine & 0.88 & Germanium Source	& 0.67\\
Gel Composition & 0.57 & Phosphorus Source	& 0.76\\
Silicon Source & 0.57 & Template Agent	& 0.43 \\
Aluminum Source & 0.53 & Molecular Sieve Structure Information	& 0.79\\
\hline
\end{tabular}
\end{table}
We structured our experimental methodology into four distinct phases. In each phase, the data was partially split in ratios of 1/3, 2/3, and 1, allowing us to systematically assess the impact of varying data volumes on model accuracy. This approach validates the hypothesis that increased data volume correlates with improved model accuracy. Additionally, to ensure the robustness of our results, we employed a ten-fold cross-validation technique for each experimental set.

\begin{figure}
    \centering
    \includegraphics[width=0.9\textwidth]{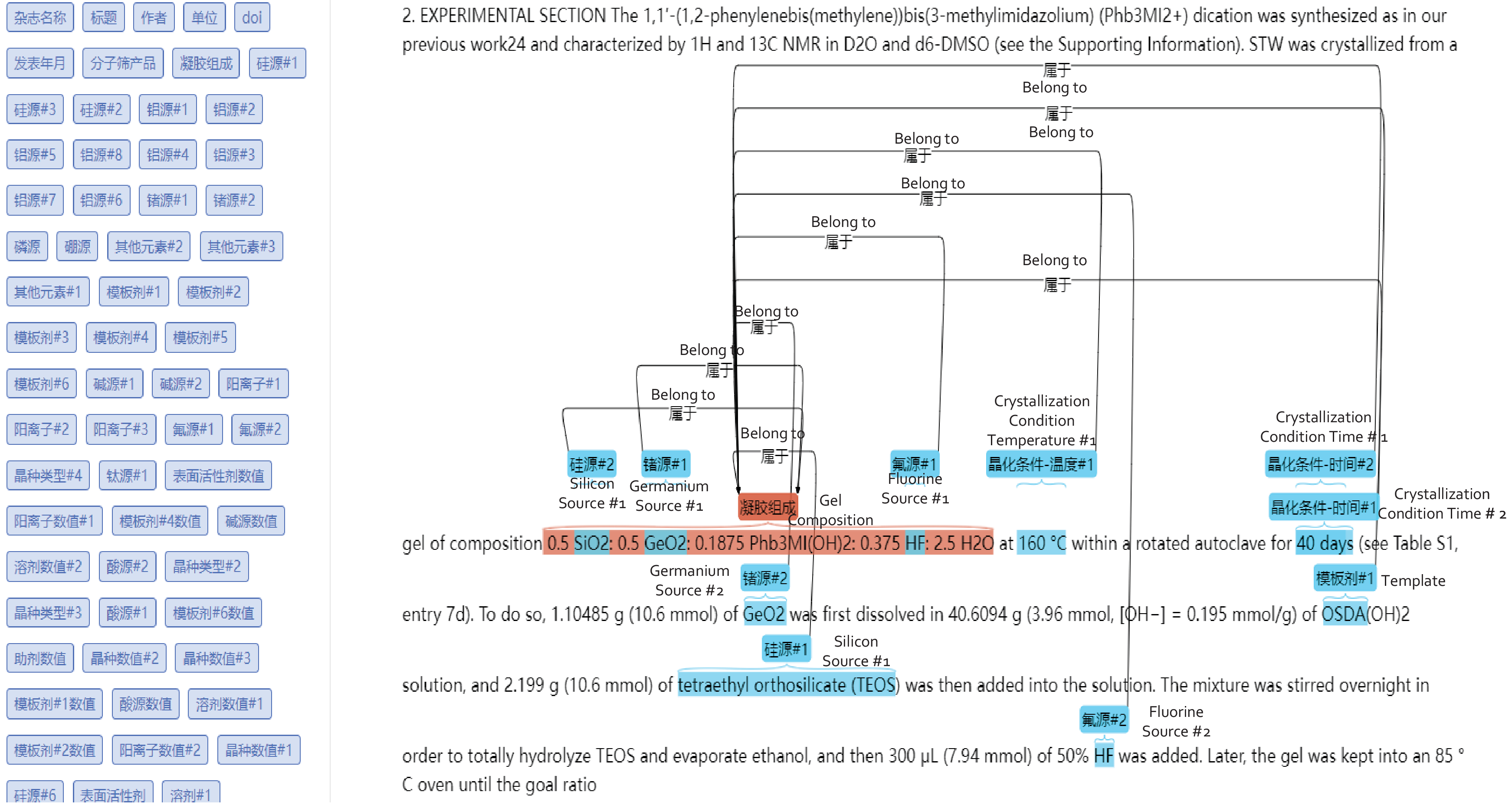}
    \caption{The display exhibits the outcomes from AutoIE. Situated on the left-hand side of the figure, one can observe an array of prevalent labels pertinent to the synthesis of molecular sieves, including but not limited to, 'Author', 'Title', 'Unit', and 'Silicon Source'.} 
    \label{fig3}
\end{figure}

The outcomes of these experiments are promising. In the molecular sieve synthesis document information extraction model framework, we observed an average accuracy of 90\% in public field extraction and 69\% in private field extraction. The overall average accuracy across all fields stands at 78\% (details shown in Table \ref{tab2}). These results highlight the efficacy of the SBERT model in this application and underscore the potential of transfer learning in enhancing the capabilities of models in highly specialized domains. Due to the article's length limitation, we directly conclude the speed gap between AutoIE and traditional manual extraction methods here. 
\begin{table}
\centering
\caption{Extraction speed of traditional extraction method and auto-extraction method.}\label{tab3}
\begin{tabular}{|l|l|l||l|l|l|}
\hline
File &	Tradition method	& Auto method & File &	Tradition method	& Auto method\\
\hline
GY-1	& 16min20s	& 3min 04s & GY-11	& 16min35s	& 3min 35s\\
GY-2	& 13min28s	& 4min 54s & GY-10	& 14min50s	& 4min 23s\\
GY-3	& 14min07s	& 3min 45s & GY-9	& 13min07s	& 3min 52s\\
GY-4	& 13min45s	& 4min 34s & GY-8	& 17min02s	& 3min 1s\\
GY-5	& 11min45s	& 3min 23s & GY-7	& 14min50s	& 2min 56s\\
GY-6	& 16min28s	& 3min 25s & -- & -- & --\\

\hline
\end{tabular}
\end{table}
In addition, we also compare the speed between the traditional extraction method and the auto-extraction method based on our framework. The test results (shown in \ref{tab3}) of the third-party trial users on the framework show that compared with the original method of molecular sieve literature information extraction, the model framework of molecular sieve synthesis literature information extraction can improve the efficiency of molecular sieve literature information extraction by more than three times. In addition, Fig. \ref{fig3} shows the final result of our framework.

\section{Conclusion}
In this research, we developed an innovative framework for extracting key information from scientific papers, encompassing four main components: MFFAPD, AFBRSC, STECIM, and OLPTMS. This framework is further augmented by introducing a relational entity joint extraction algorithm, SBERT, based on SpERT. Applying this comprehensive framework in molecular sieve synthesis literature has demonstrated its effectiveness, showcasing the potential of our approach in facilitating advanced information extraction.
In the future, our research will explore integrating general-purpose large language models, such as ChatGPT, into our existing framework. This integration aims to enhance the accuracy and efficiency of key information extraction, particularly in domain-specific scientific contexts. Our objective is to leverage the evolving capabilities of AI and machine learning to refine and expand the scope of automated information extraction in specialized scientific fields.

\end{document}